\documentclass[aip,jcp,reprint,
superscriptaddress,
footinbib,twocolumn,longbibliography]{revtex4-1}
\usepackage{graphicx}
\usepackage{amsmath}
\usepackage{amssymb} %
\usepackage{braket}
\usepackage{float}
\usepackage{bbold}
\usepackage{color}
\usepackage{bm}
\usepackage{soul} %
\usepackage{subcaption}
\usepackage{comment}
\usepackage{adjustbox}
\usepackage{placeins}
\usepackage{hyperref}
\hypersetup{colorlinks=true,linkcolor=blue,citecolor=blue,urlcolor=blue}

\newcommand{\blue}[1]{{\color{blue}{#1}}{}}
\newcommand{\eu}{\mathrm{e}^}
\newcommand{\Tr}{\text{Tr}}
\newcommand{\tr}{\text{Tr}}

\newcommand{\betaN}{\beta_{N}}
\newcommand{\F}{\mathcal{\hat{F}}}
\newcommand{\Proj}{\mathcal{\hat{P}}}
\newcommand{\Hop}{\mathcal{\hat{H}}}
\newcommand{\Fn}{\mathcal{\hat{F}}_\mathrm{n}^}
\newcommand{\Pn}{\mathcal{\hat{P}}_\mathrm{n}^}

\newcommand{\Pe}{\mathcal{\hat{P}}_\mathrm{e}^}
\newcommand{\Fe}{\F_\mathrm{e}^}
\newcommand{\iu}{\ensuremath{\mathrm{i}}}
\newcommand{\half}{\frac{1}{2}}

\newcommand{\rmd}{\mathrm{d}}
\newcommand{\der}[3][]{\frac{\rmd^{#1}{#2}}{\rmd{#3}^{#1}}}
\newcommand{\ders}[3]{\frac{\rmd^2{#1}}{\rmd{#2}\rmd{#3}}}

\begin{document}

\title{Nonadiabatic ring-polymer instanton rate theory: a generalised dividing-surface approach}
\author{Rhiannon A. Zarotiadis}
\email{rhiannon.zarotiadis@nyu.edu}
\affiliation{Department of Chemistry and Applied Biosciences, ETH Z\"{u}rich, 8093 Z\"{u}rich, Switzerland}
\affiliation{Simons Center for Computational Physical Chemistry, New York University, New York, NY 10003, USA}
\affiliation{Department of Chemistry, New York University, New York, NY 10003, USA}
\author{Joseph E. Lawrence}
\affiliation{Department of Chemistry and Applied Biosciences, ETH Z\"{u}rich, 8093 Z\"{u}rich, Switzerland}
\affiliation{Simons Center for Computational Physical Chemistry, New York University, New York, NY 10003, USA}
\affiliation{Department of Chemistry, New York University, New York, NY 10003, USA}
\author{Jeremy O. Richardson}
\email{jeremy.richardson@phys.chem.ethz.ch}
\affiliation{Department of Chemistry and Applied Biosciences, ETH Z\"{u}rich, 8093 Z\"{u}rich, Switzerland}
\date{\today}

\begin{abstract}
Constructing an accurate approximation to nonadiabatic rate theory which is valid for arbitrary values of the electronic coupling has been a long-standing challenge in theoretical chemistry.
Ring-polymer instanton theories offer a very promising approach to solve this problem, since they can be rigorously derived using semiclassical approximations and can capture nuclear quantum effects such as tunnelling and zero-point energy at a cost similar to that of a classical calculation.
A successful instanton rate theory already exists within the Born--Oppenheimer approximation, for which the optimal tunnelling pathway is located on a single adiabatic surface.
A related instanton theory has also been developed for nonadiabatic reactions using two weakly-coupled diabatic surfaces within the
framework of Fermi's golden rule. %
However, many chemical reactions do not satisfy the conditions of either limit. %
By employing a tunable dividing surface which measures the flux both along nuclear coordinates
 as well as between electronic states,
we develop a generalised nonadiabatic instanton rate theory that bridges between these two limits.
The resulting theory approximates the quantum-mechanically exact rates well for the systems studied and,
in addition, offers a novel mechanistic perspective on nonadiabatic reactions.
\end{abstract}

\maketitle

\section{Introduction}
One of the most important goals %
of modern theoretical chemistry is to predict the rate of chemical reactions and to discover their underlying mechanisms.
In many cases, the Born--Oppenheimer (BO) approximation is valid, such that the reaction is well described by a single adiabatic potential energy surface.
For chemical reactions in the BO limit, semiclassical instanton theory\cite{Miller1975semiclassical} has become an established method for calculating rates including nuclear quantum effects such as tunnelling and zero-point energy.\cite{Perspective}
It has been applied successfully in a wide range of systems from gas-phase reactions and hydrogen-bond rearrangements in water clusters to catalysis, surface processes and hydrogen abstraction in enzymes.\cite{Andersson2009Hmethane,GPR,*Muonium,Meisner2016water,hexamerprism,porphycene,TEMPO,methanol,hexamerprism,dimersurf,Hgraphene,Asgeirsson2018instanton,Lamberts2017ice,Rommel2012enzyme}
Instanton theory is rigorously derived within the path-integral description of quantum mechanics, which explains its accuracy in the deep-tunnelling regime.\cite{Miller1975semiclassical,Althorpe2011ImF,AdiabaticGreens,InstReview,Lawrence2024uniform}
The practical implementation of instanton theory is facilitated by a ``ring-polymer'' discretisation, with which the optimal tunnelling path (the instanton) can be found numerically.\cite{RPInst,Andersson2009Hmethane,Rommel2011locating,InstReview}
The BO approximation is not valid in certain types of chemical reaction, most famously for proton-coupled and pure electron-transfer reactions.\cite{Ishikita2007,Marcus1985review}
Fortunately, a semiclassical instanton theory can also be derived for reactions in the limit of weak diabatic coupling, where Fermi's golden rule (GR) holds.
\cite{GoldenGreens,GoldenRPI,inverted,PhilTransA,GRperspective}
This method has been applied successfully to study charge-transfer and spin-crossover reactions in molecular systems.\cite{Heller2021Thiophosgene, nitrene, CIinst, oxygen, carbenes}
There are, however, many reactions that are neither well described by a single adiabatic surface nor located in the golden-rule regime, which makes finding a general semiclassical nonadiabatic rate theory one of the largest open challenges in rate theory.\cite{Wigner1938TST, Lawrence2020rates, PhenolPhenoxylAdiabaticity, NonadiabaticOpenChallenge}
Such a theory should reduce to established theories in the BO and GR limits and crucially to be applicable to systems in an intermediate regime.
There have already been a number of attempts to extend instanton theory to capture nonadiabatic phenomena.
The earliest of these attempts was made by Cao and Voth. \cite{Cao1995nonadiabatic,Cao1997nonadiabatic}
After problems were identified, it was subsequently modified by Schwieters and Voth,\cite{Schwieters1998diabatic, Schwieters1999diabatic} and the revised theory was picked up in a recent publication.\cite{Ranya2020MF-MVRPI}
However, as we show in Ref.~\onlinecite{NImF}, this ``mean-field ring-polymer instanton'' (MFRPI) theory can break down in the GR limit.
There have been many attempts to take an existing path-integral method which was originally designed for %
the BO regime
and turn it into a more general nonadiabatic theory.\cite{Menzeleev2014kinetic, Kretchmer2016KCRPMD, Kretchmer2013ET, Tao2018isomorphic, Tao2019RPSH, Menzeleev2011ET, mapping, Ananth2013MVRPMD, Shushkov2012RPSH, Duke2016Faraday, Chowdhury2017CSRPMD, Lawrence2019isoRPMD}
Ref.~\onlinecite{Lawrence2020NAQI} recently presented a fundamentally new approach, the nonadiabatic quantum instanton (NAQI) approximation.
Instead of extending a theory rooted in one limit, it unifies two existing rate theories from the two limits, namely Wolynes' GR theory \cite{Wolynes1987nonadiabatic} and the adiabatic projected quantum-instanton approach.\cite{QInst}
It does so by employing a generalised dividing surface which can be optimised variationally. %
In this way it recovers the respective limits up to the accuracy of the anchoring theories.
Additionally, by unification rather than interpolation\cite{Lawrence2019ET} it allows one to extract mechanistic insight.
The NAQI perspective thus provides an ideal starting point for our nonadiabatic instanton theory.
Instead of following the approach of Ref.~\onlinecite{Lawrence2020NAQI}, which suggests statistical sampling to obtain the integral over nuclear coordinates, our nonadiabatic instanton rate theory will evaluate the integrals over space and time via the steepest-descent approximation.\cite{BenderBook}
We first introduce the exact expression for the rate in terms of the flux correlation function in Sec.~\ref{sec:exacttheory}\@.
We then discuss the steepest-descent approximation and other mathematical tools relevant for the generalised nonadiabatic rate expression in Sec.~\ref{sec:instanton}.
The key step in developing our nonadiabatic instanton theory is performed in Sec.~\ref{sec:semicl_geninst}, where we take the semiclassical approximation to the exact flux correlation function.
In Sec.~\ref{sec:tau_opt_geninst}, we describe how to evaluate the integral over time.
Finally the variational optimisation of the generalised dividing surface is discussed in Sec.~\ref{sec:alpha_opt_geninst} to obtain our final nonadiabatic instanton expression.
We compare the predictions to the exact rate and results of limiting instanton rate theories in Sec.~\ref{sec:geninst_rates} and conclude with Sec.~\ref{sec:geninst_concout}.

\section{Generalised flux correlation function}
\label{sec:exacttheory}
\begin{figure}
\centering
\includegraphics[width=.6\columnwidth]{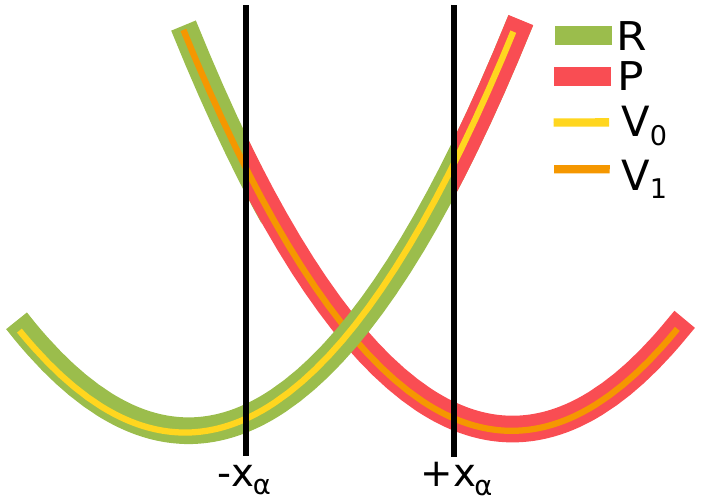}
\captionsetup{justification=raggedright, singlelinecheck=false}
\caption{{Schematic of the generalised reactant and product projections. The diabatic potentials are shown in yellow and orange. The green shading represents reactants while the red indicates the products. The two dividing surfaces (or points in this one-dimensional, symmetric system) at $\pm x_\alpha$ are shown in black.}}
\label{fig:RP_cartoon}
\end{figure}

For simplicity, we introduce our theory using a one-dimensional system but an extension to multiple dimensions can be achieved in analogy to other instanton theories.\cite{InstReview}
We define the system in the diabatic representation with the Hamiltonian  
\begin{equation}
\Hop = \frac{\hat{p}^2}{2m} + \mathbf{V}(x), \quad
\mathbf{V}(x) =
\begin{pmatrix}
V_0(x) & \Delta \\
\Delta & V_1(x)
\end{pmatrix},
\label{eq:hamiltonian}
\end{equation}
where $m$ is the nuclear mass, $\hat{p}$ is the momentum along the coordinate $x$, and the potentials corresponding to diabatic states $\ket{0}$ and $\ket{1}$ are given by $V_0(x)$ and $V_1(x)$, which are coupled by $\Delta$.
We have chosen a coordinate-independent diabatic coupling but an extension to a coordinate-dependent coupling is trivial.\footnote{The generalisation to a coordinate-dependent diabatic coupling makes minor modifications to our nonadiabatic instanton theory at two points. Firstly, it requires the evaluation of the diabatic coupling at each ring-polymer bead when optimising the tunnelling pathway. Secondly, it appears in the electronic flux operator, where it takes the value from the bead to which this operator is applied}
The derivation of a generalised flux correlation function as presented here follows Ref.~\onlinecite{Lawrence2020NAQI} closely.
In order to define the rate of a chemical reaction, one first needs to define reactants and products (R and P). %
This is achieved via the projection operators $\hat{\mathcal{P}}_\mathrm{R}$ and $\hat{\mathcal{P}}_\mathrm{P}$. %
For a given product projection $\hat{\mathcal{P}}_\mathrm{P}$, the reactant projection is defined as $\hat{\mathcal{P}}_\mathrm{R} = 1 - \hat{\mathcal{P}}_\mathrm{P}$.
While the specific details of the projection operators do not affect the rate when evaluated exactly using quantum mechanics, it is important to make an optimal choice when working with approximate theories to minimise the error.
In the BO limit, the most useful definition of a product is given by a dividing surface in nuclear position space.
The corresponding product projection, which is also used in the derivation of BO instanton theory,\cite{InstReview} is given by
\begin{equation}
\Proj_\mathrm{P}^\text{BO} = \Theta{(\hat{x})},
\label{eq:Pp_BO} 
\end{equation}
where $\Theta{({x})}$ is the Heaviside step function.
Here for simplicity we consider a system that is symmetric about $x=0$, such that we need not introduce additional parameters to define the location of the optimal dividing surface.  
\footnote{The extension to asymmetric systems can be performed by introducing an additional displacement parameter or equivalently by shifting the coordinates such that an appropriate dividing surface is located at the origin.}  %
In contrast, in the GR limit it is more natural to use a projection onto the electronic state as given by 
\begin{align}
\Proj_\mathrm{P}^\text{GR} &= \ket{1}\bra{1} = \Theta(\ket{1}\bra{1} - \ket{0}\bra{0}) = \Theta(-\hat{\sigma}_z).  
\label{eq:Pp_GR} 
\end{align}
This is the electronic-state projection employed in the derivation of GR instanton theory.\cite{GoldenGreens,PhilTransA}
Drawing inspiration from the observation that one can express both product projections using Heaviside step functions [Eqs.~\eqref{eq:Pp_BO} and~\eqref{eq:Pp_GR}], Ref.~\onlinecite{Lawrence2020NAQI} suggested introducing a generalised dividing surface as a linear combination of their arguments with a scaling parameter $x_\alpha=\tan(\alpha)$, where $\alpha \in [0, \pi/2)$.
This parameter allows one to tune the definition of reactants and products,
in order to minimise the error made by a short-time approximation to the rate.
This is similar in spirit to what is done in variational transition-state theory (VTST),\cite{VTST} although the details (discussed in Sec.~\ref{sec:alpha_opt_geninst}) differ somewhat due to the quantum nature of the problem.
The resulting generalised product projection can then be written as\cite{Lawrence2020NAQI}
\begin{align}
    \Proj_\mathrm{P}(\alpha) &= \Theta(\hat{x} - x_\alpha \hat{\sigma}_z) \nonumber \\
    &= \Theta\left(\hat{x} - x_\alpha\right)\ket{0}\bra{0} + \Theta\left(\hat{x} + x_\alpha\right)\ket{1}\bra{1} \nonumber \\
    &= \Pn{0}(\alpha)\Pe{0} + \Pn{1}(\alpha)\Pe{1}  \nonumber \\
    &= \Proj^0_\mathrm{P}(\alpha) + \Proj^1_\mathrm{P}(\alpha)
    \label{eq:general_Pp}.
\end{align}
It is comprised of direct products of projections
in nuclear position space $\Proj^{\phi}_\mathrm{n}(\alpha)$
and projections
onto an electronic state $\Proj^{\phi}_\mathrm{e}$, %
defined as
\begin{subequations}
\begin{align}
\Pn{\phi}(\alpha) &= \Theta(\hat{x} - (-1)^\phi x_\alpha), \label{eq:Pn} \\
\Pe{\phi} &= \ket{\phi}\bra{\phi}, 
\end{align}
\end{subequations}
with the diabatic state label $\phi \in \{0,1\}$.
An illustration of this definition of products and reactants can be found in Fig.~\ref{fig:RP_cartoon}.
Left of the first dividing surface at $-x_\alpha$, a nuclear configuration will be considered a reactant independent of its electronic state, and similarly, a configuration located right of the second dividing surface at $+x_\alpha$ is a product.
In between the two dividing surfaces, the electronic state determines the reactant or product nature of the system.
In practice, $x_\alpha$ will be chosen variationally (similarly to VTST) as explained below. 
Note that the generalised approach recovers the limits $\Proj_\mathrm{P}^\text{BO}$ for $\alpha = 0$ and $\Proj_\mathrm{P}^\text{GR}$ for $\alpha \rightarrow \pi/2$.
The rate is measured by the flux of particles from reactant to product and the flux operator is formally defined as the time derivative of the product projection operator, 
\begin{equation}
    \F(\alpha) = \frac{\mathrm{i}}{\hbar}[\Hop, \Proj_\mathrm{P}(\alpha)].
    \label{eq:flux_from_P}
\end{equation}
Evaluating the commutator shows that the flux operator can also be split into two terms
\begin{equation}
\F(\alpha) = \F_\text{n}(\alpha) + \F_\text{e}(\alpha),
\label{eq:general_F}
\end{equation}
which are given by
\begin{subequations} \label{eq:fluxexpansion}
\begin{align}
\F_\text{n}(\alpha) &= \Fn{0}(\alpha) \Pe{0} + \Fn{1}(\alpha) \Pe{1}, \label{eq:nfluxexpansion} \\
\F_\text{e}(\alpha) &= \Fe{0} \Pn{0}(\alpha) + \Fe{1} \Pn{1} (\alpha). 
\label{eq:efluxexpansion}
\end{align}
\end{subequations}
These operators are thus also made up of two terms, one for each electronic state $\phi$,
\begin{subequations}
\begin{align}
\Fn{\phi}(\alpha) &= \frac{1}{2m}\big[\hat{p} \, \delta(\hat{x} - (-1)^\phi x_\alpha)
+ \delta(\hat{x} - (-1)^\phi x_\alpha) \, \hat{p} \big], \label{eq:nucflux} \\
\Fe{\phi} &= \frac{\mathrm{i} \Delta}{\hbar} \left[\ket{\phi}\bra{\bar{\phi}} - \ket{\bar{\phi}}\bra{\phi}\right],
\label{eq:ele_resolvedF}
\end{align}
\end{subequations}
where $\bar{\phi}\equiv1-\phi$ indicates the other diabatic state.
In the nuclear flux operator given in Equ.~\eqref{eq:nucflux} the Dirac $\delta$-function forces the flux to be measured at one of the dividing surfaces.
The electronic flux operator, on the other hand, measures the change in population from one electronic state to the other.
In analogy to the product projection, the generalised flux operator reduces to the BO flux operator in the limit $\alpha=0$ and to the GR flux operator in the limit $\alpha \rightarrow \pi/2$.
The exact quantum-mechanical rate constant is then defined as \cite{Miller1974QTST,Miller1983rate}
\begin{equation}
k Z_\text{R} = \int_{-\infty}^{\infty} \text{d$t$ } C(t; \alpha),
\label{eq:exact_rate_time_integral}
\end{equation}
where $Z_\text{R}$ is the reactant partition function and the flux correlation function is given by
\begin{align}
C(t; \alpha) &= \frac{1}{2}\Tr\big[\F(\alpha) \eu{-\Hop (\tau - \iu t)/\hbar} 
\F(\alpha) \eu{-\Hop (\beta \hbar - \tau + \iu t)/\hbar}\big].
\label{eq:exactcff}
\end{align}
Here we use complex time with the inverse temperature $\beta= 1/k_\mathrm{B} T$ and the trace is taken over both nuclear and electronic coordinates.
The exact rate is formally independent of the imaginary-time parameter $\tau$ although a common choice is the symmetrised flux correlation function with $\tau = \beta \hbar/2$.\cite{Miller1983rate}
It is also formally independent of the choice of the product projection and therefore the $\alpha$-parameter.
Recent work has shown that a closely related flux correlation function is in fact better suited for a steepest-descent approximation.\cite{QInst}
The projected flux correlation function $c(t; \alpha)$ carries additional projections $\Proj_\mathrm{R}(\alpha)$ and $\Proj_\mathrm{P}(\alpha)$ onto the reactant and product states inside the propagators
\footnote{The explicit reactant/product projection halves the contribution to the rate, such that the factor of $\frac{1}{2}$ in Equ.~\eqref{eq:exactcff} is no longer required.}
\begin{align}
& c(t; \alpha) = \Tr\big[\F(\alpha) \, \eu{-\Hop(\tau - \iu t)/2\hbar} \, \Proj_\mathrm{R}(\alpha) \, \eu{-\Hop(\tau - \iu t)/2\hbar} \nonumber \\
&\times\F(\alpha) \, \eu{-\Hop(\beta\hbar- \tau + \iu t)/2\hbar} \, \Proj_\mathrm{P}(\alpha) \, \eu{-\Hop(\beta\hbar- \tau + \iu t)/2\hbar} \big].
\label{eq:cffRP}
\end{align}
It was shown in Ref.~\onlinecite{QInst} that this replacement does not affect the leading-order term in the semiclassical analysis and
thus Equ.~\eqref{eq:exact_rate_time_integral} is assumed to hold with $c(t; \alpha)$ to a good approximation instead of $C(t;\alpha)$.
The generalised flux operators given in Equ.~\eqref{eq:general_F} can be expanded out in the flux correlation function such that
\begin{equation}
c(t; \alpha) = c_\text{nn}(t; \alpha) + c_\text{ne}(t; \alpha) + c_\text{en}(t; \alpha) + c_\text{ee}(t; \alpha). 
\label{eq:ff_expansion}
\end{equation}
Each of the contributions are defined as 
\begin{align}
& c_{\gamma'\gamma''}(t; \alpha) = \Tr\big[\F_{\gamma'} \, \eu{-\Hop (\tau - \iu t)/2 \hbar} \, \Proj_\mathrm{R}  \,\eu{-\Hop (\tau - \iu t)/2 \hbar} \nonumber \\
&\hspace{1.5cm} \times \F_{\gamma''} \,
 \eu{-\Hop(\beta\hbar- \tau + \iu t)/2\hbar} \, \Proj_\mathrm{P} \, \eu{-\Hop(\beta\hbar- \tau + \iu t)/2 \hbar} \big],
\end{align}
where we drop the explicit $\alpha$ dependence in the operators for readability.
The subscripts $\gamma'\gamma''$ specify the nuclear--nuclear (nn), nuclear--electronic (ne), electronic--nuclear (en) and electronic--electronic (ee) terms.
\section{Steepest-descent approximation to path integrals}
\label{sec:instanton}

Exact evaluation of the quantum trace becomes numerically infeasible for systems with more than a few atoms.
Instanton theory overcomes this difficulty by evaluating the path-integral representation of the rate semiclassically, using steepest descent integration.
Here we carry out this analysis in two stages,  
first approximating the correlation functions at time $t=0$ by steepest descent
and leaving the discussion of the integral over time to Sec.~\ref{sec:tau_opt_geninst}\@.
In the following, we introduce the key mathematical concepts needed to perform the semiclassical analysis of the generalised flux correlation function.
A number of different types of integrals are required and we will explain how to compute each one in turn.

While path integrals are normally applied to systems with a scalar potential energy (such as in the BO limit), the concept can also be extended to nonadiabatic processes with two or more states,\cite{Alexander2001diabatic,Cao1997nonadiabatic,
Schwieters1998diabatic, Schwieters1999diabatic, Ranya2020MF-MVRPI, TimMasters, NRPMDChapter} where we consider the propagation not only in nuclear position space but also on the two diabatic states.
The quantum-mechanical partition function %
is given by
\begin{align}
Z &= \Tr\big[\eu{-\beta\Hop} \big].
\end{align}
The imaginary-time evolution here takes place under the diabatic Hamiltonian $\Hop$ as introduced in Equ.~\eqref{eq:hamiltonian}.
According to Feynman's path-integral formalism, this partition function can be represented by an integral over infinitely many paths.\cite{Feynman}
Its treatment is greatly simplified by discretisation of the propagator into $N$ imaginary-time slices, with corresponding inverse temperature $\betaN = \beta/N$:
\begin{align}
Z &= \lim_{N\to\infty} \Lambda^{-N} \int \text{d$\mathbf{x}$} \  \eu{-S(\mathbf{x})/\hbar} ,\label{eq:discret_part}
\end{align}
where $\Lambda = \sqrt{{2\pi\betaN\hbar^2}/{m}}$. %
The effective action is a function of the ring-polymer beads $\mathbf{x}=\{x_0,\dots,x_{N-1}\}$ and is defined by
\begin{align}
S(\mathbf{x}) = S_\text{free}(\mathbf{x}) - \hbar \ln\Big(\tr[\mathbf{M}_0 \mathbf{M}_1 \dots \mathbf{M}_{N-1}]\Big),
\label{eq:effaction}
\end{align}
with the kinetic part captured by harmonic springs connecting the beads
\begin{equation}
S_\text{free}(\mathbf{x}) = \sum_{i=1}^N \frac{m}{2\betaN \hbar} |x_i -x_{i-1}|^2
\end{equation}
(with cyclic boundary conditions $x_N\equiv x_0$) and the potential part by the matrix exponentials
\begin{equation}
\mathbf{M}_i = \mathrm{e}^{-\betaN \mathbf{V}(x_i)}.
\end{equation}
The representation becomes exact in the limit of $N\rightarrow\infty$, where the product of matrix exponentials becomes equivalent to the solution of the imaginary-time electronic Schr\"odinger equation in a basis of the two diabatic states.
From here on the large-$N$ limit will be taken to be implicit.

We will evaluate the path integral by the method of steepest descent,\cite{BenderBook}
\begin{align}
Z &\sim \Lambda^{-N} \sqrt{\frac{(2\pi\hbar)^{N}}{\text{det}_N\nabla^2 S}}\,\eu{-S(\tilde{\mathbf{x}})/\hbar},
\label{eq:sc_part}
\end{align}
where the determinant of the Hessian of the action $S$ is obtained from all $N$ rows and columns.
This asymptotic approximation is valid in the $\hbar\rightarrow0$ limit and is thus called a semiclassical approximation.
Instead of the infinite sum over paths, the semiclassical approximation relies only on one dominant trajectory, $\tilde{\mathbf{x}}$, identified as the path of minimal action.
From Sec.~\ref{sec:exacttheory} we see that partition functions are not sufficient, but that we are required to evaluate terms in the flux correlation function which involve a Dirac $\delta$-function.
An example is the (unnormalised) probability density $P_\delta$ for finding the path at $x_\alpha$ at imaginary time $\tau$ %
\begin{align}
P_\delta(x_\alpha) &= \Tr\big[\eu{-\tau \Hop/\hbar} \, \delta(\hat{x} - x_\alpha) \, \eu{-(\beta\hbar-\tau)\Hop/\hbar} \big].
\end{align}
Note that the trace is cyclic and thus the particular splitting of the Boltzmann operator is formally unnecessary; it is nonetheless introduced to make a clear connection to more complicated expressions encountered in the following.
Here, the propagator can be represented by the ring-polymer discretisation
\begin{align}
P_\delta(x_\alpha) &= \Lambda^{-N} \int  \mathrm{d}\mathbf{x} \ \eu{-S(\mathbf{x})/\hbar} \, \delta(x_{N_\tau} - x_\alpha),
\end{align}
where the bead with index $N_\tau=\lfloor N\tau/\beta\hbar \rfloor$ is constrained to the dividing surface.
By first evaluating the integral over $x_{N_\tau}$ exactly, and then performing steepest-descent integration in the remaining degrees of freedom, we obtain the semiclassical approximation
\begin{align}
P_\delta(x_\alpha) &\sim \Lambda^{-N} \sqrt{\frac{(2\pi\hbar)^{N-1}}{\text{det}_{N-1}\nabla^2 S}}\,\eu{-S(\tilde{\mathbf{x}})/\hbar}.
\label{eq:sc_PDirac}
\end{align}
Here, the stationary path, $\tilde{\mathbf{x}}$, is determined by fixing $\tilde{x}_{N_\tau}=x_\alpha$ and then minimising the action with respect to all other beads.
Finally, note that the row and column corresponding to the $x_{N_\tau}$ coordinate are removed from the Hessian before taking the determinant over the remaining $N-1$ degrees of freedom.
It is easy to extend this approach to treat multiple $\delta$-functions and/or Heaviside step functions as explained in Appendix~\ref{app:mathematical_tricks}.

\begin{figure}[H]
\centering
\includegraphics[width=1\columnwidth]{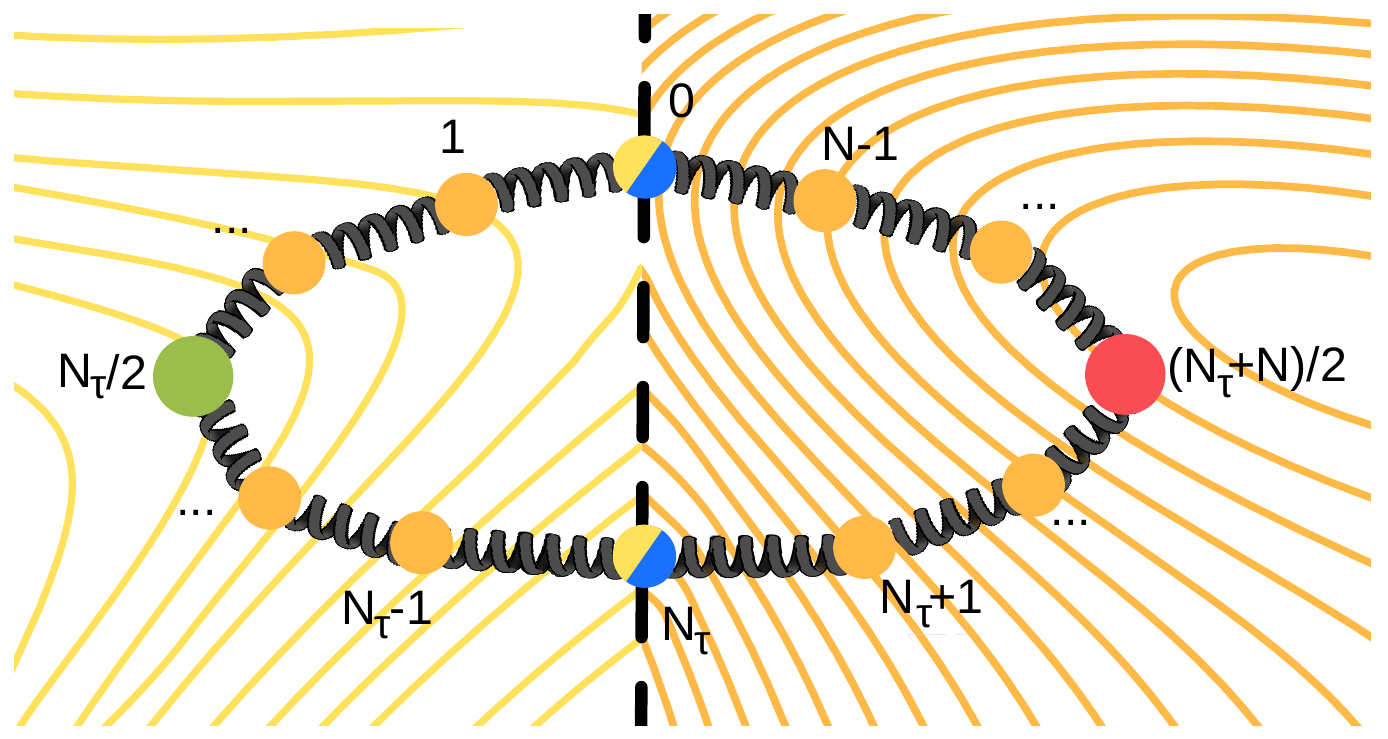}
\captionsetup{justification=raggedright, singlelinecheck=false}
\caption{{Cartoon representation of the ring polymer corresponding to the generalised effective action given in Equ.~\eqref{eq:effaction_geninst}.
The reactant projection is shown in green and the product projection in red, corresponding to the colours used in Fig.~\ref{fig:RP_cartoon}.
The flux operators may be electronic (yellow) or nuclear (blue) but either way introduce an electronic component [Equ.~\eqref{eq:fluxexpansion}].
}}
\label{fig:ring_polymer_cartoon}
\end{figure}

In addition to functions of the nuclear positions, we are also required to include operators in electronic state space.
As a particular example we give an action of the form required for an electronic--electronic term including reactant and product projections labelled by
$\kappa',\kappa''$:
\begin{align}
S(\mathbf{x}) &= S_\text{free}(\mathbf{x}) -\hbar \ln \Big(\tr[\mathbf{M}_0^{1/2} \Fe{\phi'} \mathbf{M}_0^{1/2} \mathbf{M}_1 \cdots \nonumber \\
&\qquad \mathbf{M}_{N_\tau/2}^{1/2} \Pe{\kappa'} \mathbf{M}_{N_\tau/2}^{1/2} \cdots \mathbf{M}^{1/2}_{N_\tau} \Fe{\phi''} \mathbf{M}^{1/2}_{N_\tau} \cdots \nonumber \\
&\qquad\mathbf{M}^{1/2}_{(N_\tau+N)/2} \Pe{\kappa''} \mathbf{M}^{1/2}_{(N_\tau+N)/2} \cdots \mathbf{M}_{N-1}]\Big),
\label{eq:effaction_geninst}
\end{align}
where $\mathbf{M}_i^{1/2} = \mathrm{e}^{-\betaN\mathbf{V}(x_i)/2}$.
The location of the first flux operator is always set to be at bead index $0$ and the second flux operator at $N_\tau$.
The reactant and product projections are always positioned halfway between the flux operators and a cartoon representation of this ring polymer is shown in Fig.~\ref{fig:ring_polymer_cartoon}.
All projections are placed ``inside'' the beads meaning the projection is placed between two matrices of the same bead where the magnitude of the exponent is halved.
These are the key mathematical techniques required to perform the semiclassical approximations of the nonadiabatic flux correlation functions at time $t=0$ introduced in Sec.~\ref{sec:exacttheory}.

\section{Semiclassical analysis of the generalised flux correlation function}
\label{sec:semicl_geninst}
\begin{figure}
    \centering
    \includegraphics[width=.85\columnwidth]{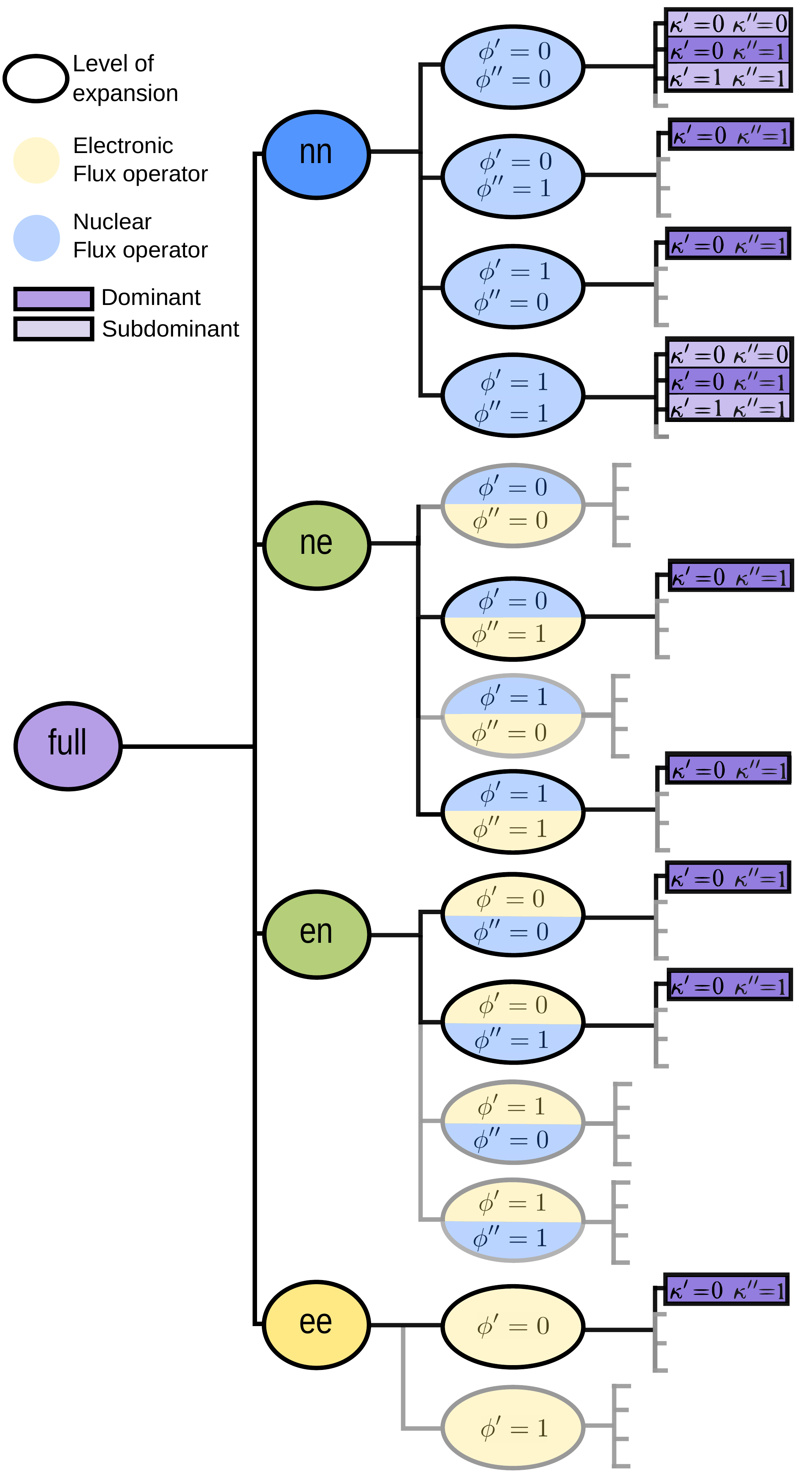}
    \captionsetup{justification=raggedright, singlelinecheck=false}
    \caption{{Flow chart of the different terms in the expansion of the flux correlation function at time $t=0$ within the semiclassical analysis.
    The boxes at the end of the chart require separate instanton optimisations.
    The ones coloured purple are dominant and the ones coloured light purple are around one order of magnitude smaller than the dominant term.
    Grey branches indicate the existence of negligible terms. %
    }}
    \label{fig:inst_tree}
\end{figure}

We have introduced the generalised flux correlation function in Sec.~\ref{sec:exacttheory}\@.
In this section, we analyse how it can be decomposed into various terms that can be approximated by semiclassical instanton pathways using the tools outlined in Sec.~\ref{sec:instanton}\@.
This will involve the expansion of the correlation function into a number of terms, of which only a small subset will actually be required in the final calculation.
We have already seen in Sec.~\ref{sec:exacttheory} that the separation of the flux operator into a nuclear and electronic part results in four different contributions [Equ.~\eqref{eq:ff_expansion}]. 
These terms form the first layer of our expansion, defining four different classes of terms,  with each flux operator introducing an index $\gamma'\in\{\rm n,e\}$ and $\gamma''\in\{\rm n,e\}$.
The first layer in the flowchart in Fig.~\ref{fig:inst_tree} depicts this expansion.

To form the second layer, each of the flux operators can be expanded out further
by introducing the indices $\phi'$ and $\phi''$.
In the case of the nuclear flux operators they simply correspond to the expansion shown in Equ.~\eqref{eq:nfluxexpansion}.
However, as we shall discuss in the following, the precise meaning of $\phi'$ and $\phi''$ may differ when expanding the electronic flux operators. %
The final layer is formed by expanding out the reactant and product projections in terms of electronic states [Equ.~\eqref{eq:general_Pp}], where 
the electronic-state projections in the reactant/product projections are denoted with $\kappa' \in \{0, 1\}$ and $\kappa'' \in \{0, 1\}$.
This is done so as to account for the effect of their nuclear projections in terms of Heaviside step functions when solving the steepest-descent integral in nuclear position space.
\FloatBarrier
Finally, for all but the electronic--electronic term, we can write the expansion in four indices to give the complete set of components of the correlation function
\begin{subequations}
\begin{align}
c_{\gamma'\gamma''}(t) %
&= \sum_{\phi'\phi''\kappa'\kappa''}^{} c^{\phi'\phi''\kappa' \kappa''}_{\gamma'\gamma''}(t),
\label{eq:general_expansion}
\end{align}
where we have dropped the explicit dependence on the $\alpha$ parameter for readability.
In the case of the electronic--electronic term, for reasons discussed in Sec.~\ref{sec:ee}, an expansion in only three indices is required such that the expression is
\begin{equation}
c_\mathrm{ee}(t)=\sum_{\phi'\kappa'\kappa''}^{} c^{\phi'\kappa' \kappa''}_\mathrm{ee}(t).
\end{equation}
\end{subequations}
It is obvious that not all of the terms will be equaly important, and in many cases, it will be valid to assume that the terms with $\kappa'=0$ and $\kappa''=1$ dominate (because the first is associated with $\mathcal{P}_\mathrm{R}$ and the second with $\mathcal{P}_\mathrm{P}$), greatly reducing the number of instantons to optimise.
Note, that we do include %
some of the other terms whenever they are not negligible, e.g.\ when calculating the nuclear--nuclear terms in the adiabatic limit.
In the flowchart of Fig.~\ref{fig:inst_tree}, we have coloured the dominant instantons dark purple and terms which are sometimes non-negligible light purple.
The subdominant terms (indicated by grey branches) are not considered in the final rate calculations simply to avoid superfluous effort.

We still have to define exactly how the various terms are split into their separate components.
Within the quantum-mechanical framework, the exact rate can be obtained regardless of how the expansion is performed.
However, for our semiclassical theory, we wish to perform a steepest-descent integration over time around the $t=0$ value, and thus it is important that each term in the flux correlation function decays quickly.
In the following, our choice of how to expand each term is guided by this principle.
\subsection{Nuclear--nuclear terms}
\begin{figure}
    \centering
    \includegraphics[width=\columnwidth]{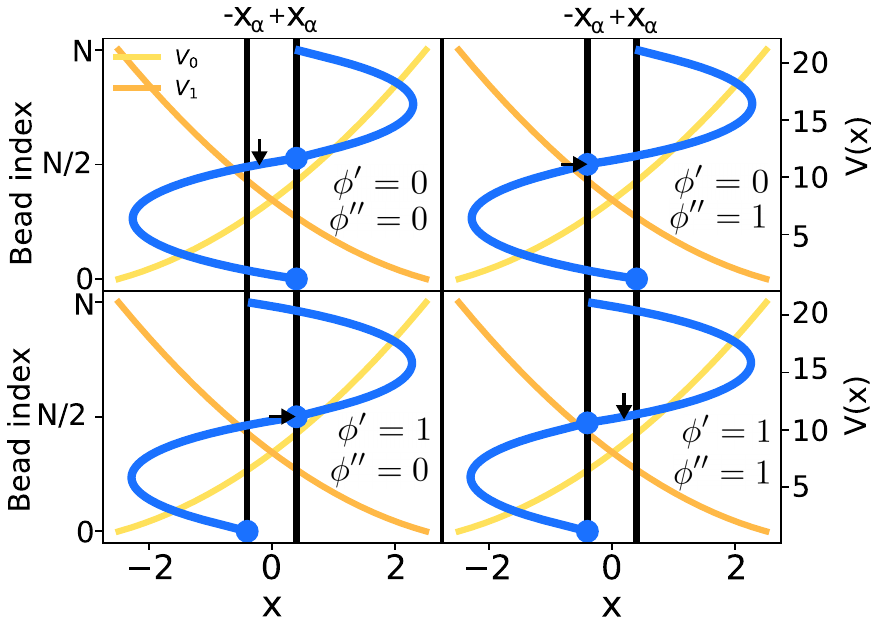}
    \captionsetup{justification=raggedright, singlelinecheck=false}
    \caption{{
    Plot of the different types of nuclear--nuclear instantons.
    The blue dots indicate the location of the flux operators according to the optimal value of the parameter $\tau$.
    The black arrows mark the halfway point of the path in imaginary time ($\tau = \beta \hbar/2$).
    This highlights the (a)symmetry of each of the instanton terms.
    The black vertical lines indicate the two generalised dividing surfaces at $\pm x_\alpha$.
    The diabatic PESs are given in yellow and orange.
    }}
    \label{fig:nn_paths}
\end{figure}

The first term we consider is the nuclear--nuclear term (coloured all blue in Fig.~\ref{fig:inst_tree}, first layer).
It includes Dirac $\delta$-functions
placed at $x=-x_\alpha$ for the nuclear flux operator $\Fn{1}(\alpha)$ and at $x=+x_\alpha$ for $\Fn{0}(\alpha)$.
Importantly, as the $\delta$-functions correspond to hard constraints on the path, 
each term with a different nuclear flux operator has to be evaluated separately within the steepest-descent approximation.
We thus fully expand out both flux operators (indicated by $\phi'$ and $\phi''$) and we also fully expand out the reactant and product projections (using the indices $\kappa'$ and $\kappa''$).
The terms in Equ.~\eqref{eq:general_expansion} for the nuclear--nuclear flux correlation function are thus defined as
\begin{align}
    c^{\phi' \phi'' \kappa' \kappa''}_\mathrm{nn}(t) = \Tr\big[\Fn{\phi'}\Pe{\phi'} \hat{K}' \Proj^{\kappa'}_\mathrm{R} \hat{K}' 
    \Fn{\phi''} \Pe{\phi''} \hat{K}'' \Proj^{\kappa''}_\mathrm{P} \hat{K}'' \big],
    \label{eq:nn_terms}
\end{align}
with each of these terms corresponding to the optimisation of a separate instanton.
Here, we have introduced a compact notation for the propagators $\hat{K}' = \eu{-\Hop (\tau - \iu t)/2\hbar}$ and $\hat{K}'' = \eu{-\Hop(\beta\hbar-\tau + \iu t)/2 \hbar}$.
The resulting nuclear--nuclear instantons are illustrated in Fig.~\ref{fig:nn_paths}, in which the nuclear flux operators (indicated by a blue dot) cause the path to be pinned to a dividing surface.
Thus, in general, 16 different types of paths have to be considered for the nuclear--nuclear contribution to the flux correlation function.
However, many of these give negligible contributions and we thus found that considering between four and eight different instantons is sufficient (as indicated in Fig.~\ref{fig:inst_tree}).
In fact, this number can be reduced even further for symmetric reactions.
\subsection{Mixed terms}
The mixed (nuclear--electronic and electronic--nuclear) terms contain both kinds of flux operators.
We begin by analysing the nuclear--electronic term.
The nuclear flux operator is expanded out according to Equ.~\eqref{eq:nfluxexpansion} into two terms indexed by $\phi'$.
The electronic flux operator does not include $\delta$-functions and we are thus more flexible about how we choose to group the terms.
In particular, we use $\Fe{0}=-\Fe{1}$ to write
$\Fe{}=\Fe{0}\Pn{0}+\Fe{1}\Pn{1}=\Fe{1}(\Pn{1}-\Pn{0})$.
Finally, we find that it is necessary to split the $\Fe{1}$ operator into its two constituent terms (labelled by $\phi''$) to give
\begin{align}
&c^{\phi' \phi'' \kappa' \kappa''}_\mathrm{ne}(t) = (-1)^{\phi''} \frac{\mathrm{i} \Delta}{\hbar} \Tr\big[\Fn{\phi'}\Pe{\phi'}\hat{K}' \Proj^{\kappa'}_\mathrm{R} \hat{K}'
\nonumber \\
&\qquad\times
\ket{\bar{\phi}''}\bra{\phi''} (\Pn{1}-\Pn{0}) \hat{K}'' \Proj^{\kappa''}_\mathrm{P}\hat{K}''\big].
\label{eq:cff_ne_sum}
\end{align}
Note, that despite the imaginary prefactor, this term is real-valued since there is another factor of $\iu$ in the nuclear flux operator.
Again, every term that is written as one trace requires one instanton optimisation. %
The operator $(\Pn{1}-\Pn{0})$ simply forces the corresponding bead ($N_\tau$ in this case) to lie between $-x_\alpha$ and $x_\alpha$.
We found that the contribution with $\kappa'=0$ and $\kappa''=1$ is dominant and the terms with alternative combinations of $\kappa'$ and $\kappa''$ can safely be neglected.
This reduces the number of instantons to consider from 16 to four.
In addition, %
considering the location of the $\ket{\bar{\phi}''}\bra{\phi''}$ operator between the reactant and product projections,
all terms with $\phi''=0$ are also subdominant further reducing the number of relevant contributions to two.

The electronic--nuclear term is treated in the same way with the order of the flux operators swapped, and the index of the dominant term with respect to the electronic flux operator is switched accordingly, too.
\subsection{Electronic--electronic terms} \label{sec:ee}
The electronic--electronic term is expanded out with respect to both electronic flux operators.
In order to obtain an expression that is well-approximated by steepest descent we found it necessary to regroup one of the electronic flux operators such that the expansion of the generalised flux correlation function is only required in terms of three indices.
Each term in this expansion is written as
\begin{align}
&c^{\phi' \kappa' \kappa''}_\mathrm{ee}(t) = \frac{\Delta^2}{\hbar^2} \Tr\big[\ket{\bar\phi'}\bra{\phi'}(\Pn{1}-\Pn{0}) \hat{K}' \Proj^{\kappa'}_\mathrm{R}\hat{K}' \nonumber \\
&\times \left(\ket{\phi'}\bra{\bar\phi'} - \ket{\bar\phi'}\bra{\phi'}\right) (\Pn{1}-\Pn{0})
\hat{K}'' \Proj^{\kappa''}_\mathrm{P}\hat{K}''\big].
\end{align}
This regrouping of terms in the expansion of the electronic--electronic flux correlation function is required because only for this 
combination do the individual components of the flux correlation function decay rapidly in real time, such that we can integrate over them
by steepest descent (which we require to ultimately obtain a rate coefficient).
In all cases investigated here, the term with $\phi' = 0$, $\kappa'=0$, $\kappa''=1$ dominates and only one instanton needs to be optimised.
This term closely resembles the instanton expression used in GR instanton rate theory.\cite{GoldenGreens}
In fact, when the dividing surface is located at large values $x_\alpha$ the generalised expression rigorously recovers GR instanton theory.

In this way, we have split the correlation function into multiple terms of type nn, ne, en and ee.
Formally, each of these terms can be evaluated numerically exactly and their sum recovers the exact quantum-mechanical projected flux correlation function from which the rate can be obtained.
However, a more computationally efficient alternative is now possible.
At $t=0$, each of these terms can be represented as an imaginary-time path-integral and evaluated using steepest-descent approximations, as outlined in Sec.~\ref{sec:instanton} and the Appendices.
This forms the main step along our route to deriving a nonadiabatic instanton theory.
\section{Steepest-descent approximation in time}
\label{sec:tau_opt_geninst}

In order to obtain the rate, we also need to evaluate the integral of the flux correlation function over time [Equ.~\eqref{eq:exact_rate_time_integral}].
Following the Born--Oppenheimer and golden-rule instanton derivations of Ref.~\citenum{InstReview}, this integral is also performed by steepest-descent around $t=0$.

While for an exact solution, the imaginary-time parameter $\tau$ may be chosen freely as long as $0 < \tau < \beta \hbar$, to evaluate the time integral by steepest descent, $\tau$ has to correspond to a maximum of the effective action, $S^{\chi}$. 
Here, each $\chi$ denotes a specific term in the expansion of the flux and projection operators (and hence an instanton), and we have grouped all indices together as $\chi = \{ \gamma' \gamma'' \phi' \phi'' \kappa' \kappa'' \}$.
It thus has to hold that $\mathrm{d} S^{\chi}/\mathrm{d}\tau = 0$, which in general requires a different value of $\tau$ for each instanton. %
Note that, even for a symmetric system, $\tau$ might not be $\beta\hbar/2$ for certain terms, although for the systems considered here it was never found to deviate far from this value (see Fig.~\ref{fig:nn_paths}).
In this work, the total derivative with respect to $\tau$ is obtained by calculating the effective action $S^{\chi}$ on a grid of points with spacing $\beta_N \hbar$ and then applying the method of finite difference.
Note, however, that it would alternatively be possible to obtain an analytic expression for these derivatives.

The final equation which defines our new nonadiabatic rate theory is given by summing over each instanton contribution according to 
\begin{align}
k Z_\text{R} &\sim \sum_{\chi}
\sqrt{2 \pi \hbar}%
\left(-\frac{\mathrm{d}^2 S^{\chi}}{\mathrm{d}\tau^2}\right)^{-1/2} c^{\chi}(0) \label{eq:finaltheory}
\end{align}
where $c^{\chi}(0)$ is the semiclassical approximation to the projected flux correlation function with the set of indices $\chi$ at time $t=0$.
The integrals over nuclear space and time have been carried out sequentially here to aid the exposition of the theory.
Although not used here, a formally equivalent rate expression can be obtained by evaluating all steepest-descent integrals simultaneously.
This is expected to be a more efficient implementation and would require an optimisation of $\mathbf{x}$ and $\tau$ together, similarly to what is done for GR instanton theory.\cite{PhilTransA} %
\section{Variational optimisation of the generalised dividing surface}
\label{sec:alpha_opt_geninst}

\begin{figure*}
    \centering
    \includegraphics[width=1\linewidth]{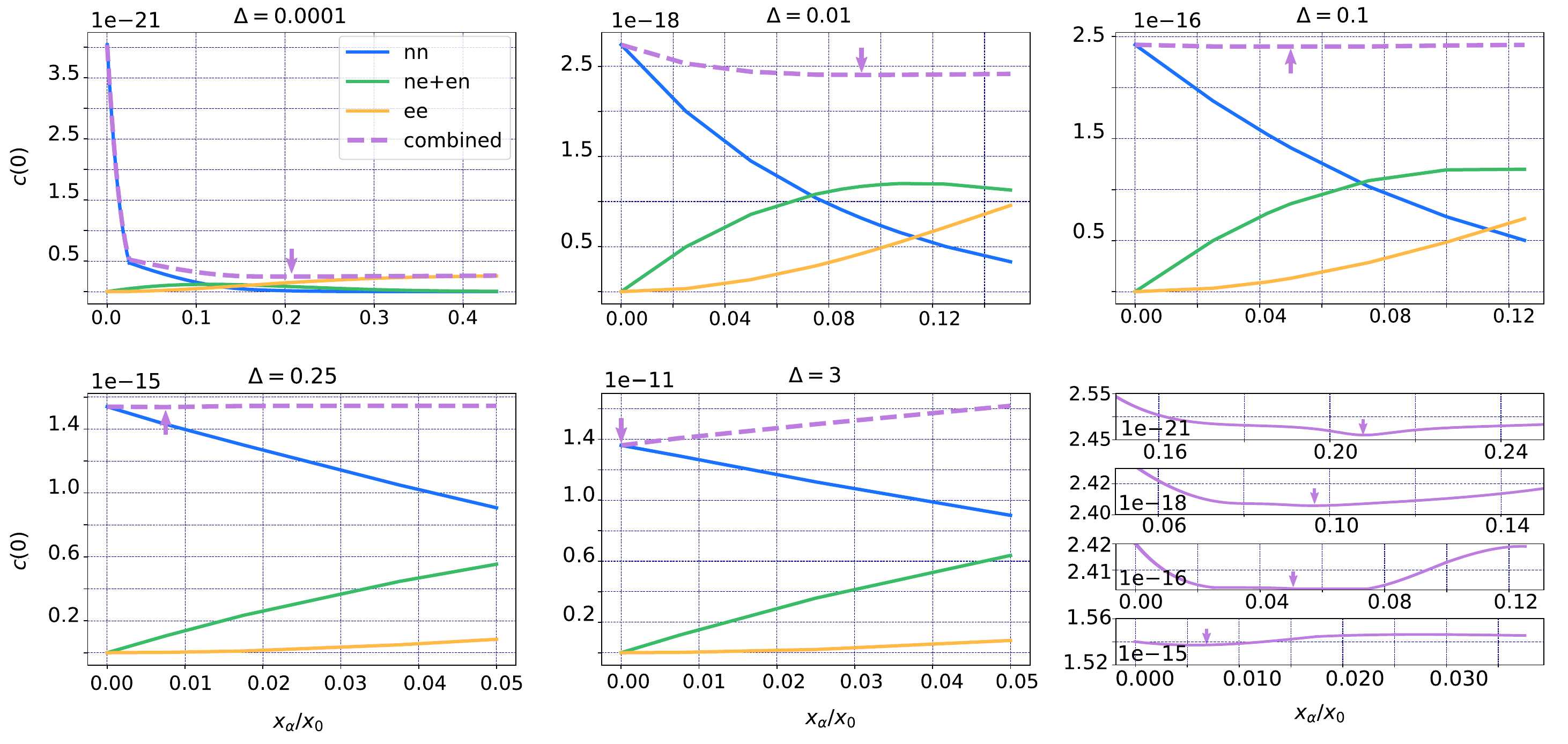}
    \captionsetup{justification=raggedright, singlelinecheck=false}
    \caption{
    Magnitude of the different contributions to the flux correlation at time $t=0$ versus the location of the dividing surface $x_\alpha$ relative to the location of the well of the symmetric harmonic oscillators $x_0$ for various values of $\Delta$ as indicated.
    The optimal $x_\alpha$ value reduces as expected when moving from a system in the GR limit to the BO limit with a large diabatic coupling. 
    The bottom, right figure shows insets of the first four smallest coupling strengths, thus highlighting the presence of minima along the combined flux correlation function.
    }
    \label{fig:sum_contributions}
\end{figure*}
In our framework, we have employed a generalised concept of a dividing surface.
Going beyond the traditional understanding of dividing surfaces used in classical TST, it not only employs a separation in nuclear coordinate space via its location $x_\alpha$ but also uses the projections onto electronic states (see Fig.~\ref{fig:RP_cartoon}) The $\alpha$ parameter is introduced to tune smoothly between the purely electronic and purely nuclear definitions of the dividing surface.
In analogy with variational transition state theory, we choose $\alpha$ so as to minimise the sum of the contributions to the flux correlation function at $t=0$ given their individual optimal values of $\tau$.\footnote{Note that this procedure differs from Ref.~\citenum{Lawrence2020NAQI}, where no separation into terms was made, and hence $\alpha$ was chosen to minimise the full correlation function with a single global value of $\tau$.}
This minimises recrossing (at least approximately), which is neglected by our theory,
and thus we expect the optimised dividing surfaces to lead to the best approximation for the rate.
Note that following Equ.~(\ref{eq:exact_rate_time_integral}), all terms in Equ.~(\ref{eq:finaltheory}) must have the same value of $\alpha$. However, the optimal choice of $\alpha$ is system- and temperature-dependent.
Within our nonadiabatic instanton rate theory, the full flux correlation function is obtained by combining the four classes of terms within the expansion Equ.~\eqref{eq:ff_expansion} as shown in the flowchart in Fig.~\ref{fig:inst_tree}.
The $\alpha$-dependence of each class is shown in Fig.~\ref{fig:sum_contributions}, for a system of two coupled harmonic oscillators with varying diabatic coupling (see Appendix~\ref{app:system} for details on the model).
In each case, the nn-contribution is non-zero at $x_\alpha=0$ and monotonically decreases to zero as $x_\alpha$ increases.
In contrast, the ee-term is zero at $x_\alpha = 0$ and increases with increasing $x_\alpha$ up to a plateau value, as can be seen most clearly in the top left panel of Fig.~\ref{fig:sum_contributions}.
The mixed terms are also zero at $x_\alpha=0$; their magnitude increases for intermediate $x_\alpha$ values and then decreases again as $x_\alpha$ becomes large.
We combine the terms together to obtain the full correlation function as a function of $\alpha$ and then locate its minimum.
For a system with small diabatic coupling (see Fig.~\ref{fig:sum_contributions}, top left), the optimal $x_\alpha$ value, which is indicated by an arrow, is large (more than 20\% of the distance to the well bottom).
The full flux correlation function decays sharply away from $x_\alpha =0$ to a plateau value at large $x_\alpha$ values.
This suggests that there is a strong desire not to put the dividing surface in the middle, but that the function is insensitive towards the optimal location as long as it is large enough.
The dividing surfaces are thus pushed far away from the barrier region and the electronic states determine reactants and products in line with our expectations for a system in the GR limit.
In addition, at the optimal $x_\alpha$ value, the ee-contribution, which is most similar to the GR instanton, dominates the flux correlation function.
This ensures that our nonadiabatic instanton rate theory recovers GR instanton theory in the $\Delta\rightarrow0$ limit. %
As the diabatic coupling is increased (see Fig.~\ref{fig:sum_contributions}, top row), the optimal $x_\alpha$ value decreases.
Nevertheless, the full flux correlation function (i.e.\ the sum of the contributions) at its optimal value of $\alpha$ still exhibits approximately $\Delta^2$-dependence, %
i.e.\ $c(0)/\Delta^2\approx2.4\times10^{-14}$ for $\Delta\lesssim0.1$, implying that it is still in the GR regime.

For larger values of $\Delta$, there is a qualitative change in the behaviour of the function (Fig.~\ref{fig:sum_contributions}, bottom row).
The function no longer decreases away from 0, but becomes a monotonically increasing function of $x_\alpha$.
This means that for the case of $\Delta=3$, the optimal value of $x_\alpha$ is 0, %
the two dividing surfaces coalesce and the flux correlation function is completely composed of the nn-contribution.
This rigorously recovers the definition of reactant and product used in BO instanton theory,
although even in this case, there may still be nonadiabatic effects arising from the propagators.
These effects slowly disappear in the BO limit, such that
the subsequent rate expression %
recovers the adiabatic instanton rate (except for a minor difference due to the way in which the reactant and product projections are applied).
The value of the optimal $\alpha$ parameter thus provides a rough indication of whether the system should be considered as close to the GR or BO limits.
Mechanistic interpretation is obtained from the various instanton tunnelling pathways and the magnitude of their contributions at the optimal $\alpha$ value.

\section{Semiclassical rate prediction}
\label{sec:geninst_rates}
The effort required to calculate our nonadiabatic instanton rates is ultimately similar to that of other nonadiabatic instanton methods.\cite{Schwieters1998diabatic, Schwieters1999diabatic, Ranya2020MF-MVRPI}
A key difference to the BO case is that two PESs and the couplings have to be calculated at each bead, and instead of one, the number of instantons to be calculated is a minimum of nine for an asymmetric system or as few as four for a symmetric system.

\begin{figure}
    \centering
    \includegraphics[width=.99\columnwidth]{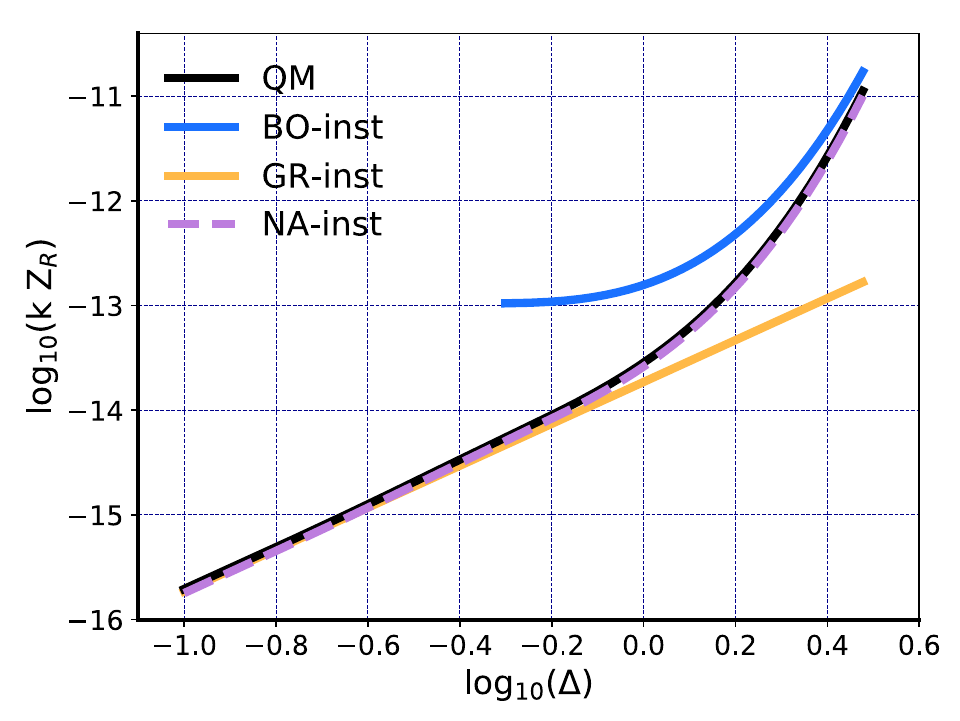}
    \captionsetup{justification=raggedright, singlelinecheck=false}
    \caption{{Logarithm of the rate versus the logarithm of the diabatic coupling $\Delta$. The exact quantum-mechanical (QM) rate is shown in black, the BO instanton prediction in blue and the GR instanton prediction in yellow.
    The prediction of the nonadiabatic instanton rate theory is shown in purple and agrees well with the benchmark over the full range of coupling strengths.
    }}
    \label{fig:Delta_rate}
\end{figure}

Figure \ref{fig:Delta_rate} shows the rates calculated by BO instanton theory,\cite{RPInst, InstReview} GR instanton theory\cite{GoldenRPI} and the new nonadiabatic instanton rate theory for the system of coupled harmonic oscillators defined in Appendix~\ref{app:system}.
This system is in the deep-tunnelling regime (see Table~\ref{tab:rates} for a comparison to classical rate predictions).
The instanton rates are calculated with $N=300$ beads, which we found to be sufficient for convergence.
Figure~\ref{fig:Delta_rate} shows that, in addition to capturing the BO (large $\Delta$) and GR (small $\Delta$) limits, as expected from our analysis, nonadiabatic instanton theory is an excellent approximation to the exact quantum-mechanical rate across the entire range of coupling strengths, including in the most interesting intermediate regime.

\section{Conclusion and Outlook}
\label{sec:geninst_concout}
In this work, we have derived
a nonadiabatic instanton rate theory %
which successfully predicts rates for the full span of diabatic coupling strengths.
There are three key concepts contributing to the success of our new theory.
Firstly, we employ a generalised definition of reactants and products.
In doing so we recover the BO-style definition for large diabatic coupling $\Delta$, where the location along the nuclear coordinate is the identifier of the reactant and the product, and the GR-style definition in the weak-coupling limit, which is determined by the diabatic state on which the path is evaluated.
This is in contrast to previous attempts at finding a generalised instanton rate theory where at most one limit could be recovered rigorously.\cite{Schwieters1998diabatic, Schwieters1999diabatic, Ranya2020MF-MVRPI, NImF}
Next, we employ a steepest-descent approximation to each term in the correlation function, each of which requires a separate instanton optimisation.
These are accompanied by a steepest-descent approximation in time to obtain an overall rate constant.
Finally, we employ the fundamental concept of transition-state theory, namely optimising the dividing surface so as to minimise the effects of recrossing.
The optimal dividing surface thus shifts as the diabatic coupling strength is changed.
Its location is not only a technical detail but gives mechanistic insight.
Comparing it to its role in the limiting rate theories allows one to gauge the adiabaticity of the reaction under study.
For this proof-of-principle study we employed a symmetric system.
The theory as presented here is however readily applicable to an asymmetric rate problem.
The reason for this lies in the separate optimisation of the imaginary time parameter $\tau$ for each contribution of the generalised rate expression.
Additionally, the introduction of the reactant and product projections ensures the correct behaviour for asymmetric systems, as shown in Ref.~\onlinecite{QInst}. 
Note that for an asymmetric system, a second parameter will be required to place the dividing surfaces independently.
An important advantage is that instanton theories are computationally efficient even in high-dimensional systems.
In fact, there have been a number of \textit{ab initio} studies for both the BO and GR limits.\cite{Heller2021Thiophosgene, nitrene, porphycene, Hgraphene, methanol, TEMPO, CIinst, oxygen, carbenes, Asgeirsson2018instanton, dimersurf}
In order to also make the nonadiabatic instanton theory applicable to molecular simulations, a multidimensional implementation of the method is necessary.
This can be derived in close analogy to the multidimensional extensions of the BO and GR instanton theories by simply working in a higher-dimensional space.
It is only the definition of reactants and products which requires additional consideration.
In the current one-dimensional formulation, the reaction is set to take place along a single nuclear (reaction) coordinate and the effect of the generalised dividing surface is applied along this coordinate.
Testing the multidimensional extension is left for future work, but in general will require a procedure akin to that of VTST.\cite{VTST}
Typically, an appropriate linear combination of the nuclear coordinates is chosen as the reaction coordinate, for instance along the imaginary normal mode at the transition state.
This simple procedure may be sufficient for our means, as from our proof-of-principle study, the quality of the nonadiabatic instanton rate prediction is expected to be relatively insensitive towards the choice of the dividing surface.
The study of quantum dynamical properties goes hand in hand with the electronic structure of the reaction. 
It is therefore important to consider the compatibility of nonadiabatic instanton rate theory with common electronic-structure methods.
At each step of the optimization, it is necessary to calculate the potential energies, gradients and diabatic couplings for each bead in the ring polymer, and once the optimization is converged, it is necessary to calculate the hessians.
Some electronic-structure methods exist which give direct access to diabatic states, \cite{Kaduk2011cDFT, Kubas2015cDFT} while for others efforts are being made to employ machine-learning algorithms in the quest for diabatic states,\cite{Parker2020methylamine,Axelrod2022ML} and work to devise quasi-diabatisation algorithms is on-going.\cite{ConicalIntersections1, Pacher1988diabatization, Neville2020diabat,diabatization}
These efforts may provide a pathway to remain in the diabatic representation for nonadiabatic rate calculations.
Alternatively, there exists no methodological reason prohibiting the recasting of the present nonadiabatic instanton theory into the adiabatic representation where adiabatic PESs from electronic-structure theory would then be more readily available.
We are thus not limited to simple models like the one we study here.
We have developed our new nonadiabatic instanton rate theory to be able to study important and inherently nonadiabatic reactions.
Proton-coupled electron transfer (PCET) reactions are an archetypal example for such nonadiabatic reactions, often located intermediate between the GR and BO limit.
PCET reactions may steer enzyme activity, be involved in photosynthesis, or promote catalysis via metal complexes.\cite{HammesSchiffer2008PCET, HammesSchiffer2004, Ishikita2007, Harshan2015}
A second highly-relevant set of reactions are those involving intersystem crossing.
They are as prevalent as PCET reactions and they are inherently nonadiabatic since they involve two electronic spin states.
Generally, it cannot be assumed that the spin--orbit coupling is small enough such as to warrant the application of Fermi's golden rule or higher-order perturbative methods originating in the GR limit.\cite{Trenins2020}
Our nonadiabatic instanton rate theory will allow us to rigorously and reliably investigate such reactions.

\section*{Acknowledgements}
RAZ and JEL were supported by Independent Postdoctoral Fellowships at the Simons Center for Computational Physical Chemistry, under a grant from the Simons Foundation (839534, MT). JEL also acknowledges support from an ETH Zurich Postdoctoral Fellowship, during which the majority of this work was conducted.
JOR acknowledges financial support from the Swiss
National Science Foundation through Project 207772.

\section*{Author contributions}
\textbf{Rhiannon Zarotiadis}: Investigation (lead); software (lead); conceptualization (supporting); methodology (equal); visualization (lead); writing -- original draft (lead); writing -- review and editing (equal).
\textbf{Joseph E. Lawrence}: Conceptualization (supporting); methodology (equal); writing -- review and editing (supporting).
\textbf{Jeremy O. Richardson}: Conceptualization (lead); methodology (equal); supervision (lead);  writing -- review and editing (equal).

\section*{Data Availability}
The data that supports the findings of this study are available within the article. 

\appendix

\section{Model system and its rate predictions}
\label{app:system}
In this proof-of-principle study we investigated a one-dimensional symmetric model consisting of two harmonic oscillators with a constant diabatic coupling $\Delta$.
The quantum-mechanically exact rate was obtained using a finite-basis representation.  In order to obtain a well-defined rate, the potentials were modified to represent scattering boundary conditions:
\begin{subequations}
\begin{align}
V_0(x) &= \begin{cases}
\frac{1}{2} m \omega^2 (x + x_0)^2 & \text{ for } x>-x_0 \\
0 & \text{ otherwise }\\
\end{cases} \nonumber\\
V_1(x) &= \begin{cases}
\frac{1}{2} m \omega^2 (x - x_0)^2 & \text{ for } x < x_0 \\
0 & \text{ otherwise }.
\end{cases} \nonumber
\end{align}
\end{subequations}
The parameters are chosen as $\hbar=1, m=1$, $\omega = 1$, $x_0 = 4$ in reduced units with the inverse temperature $\beta = 6$.

Table~\ref{tab:rates} presents the rate predictions from various methods to highlight the importance of nonadiabaticity and nuclear quantum effects.
Note that there is a small difference between the GR instanton rate and our new nonadiabatic instanton rate for small diabatic coupling strengths.
The variational optimisation of the $\alpha$ value is performed on a finite grid. The placement of the optimal value of $x_\alpha$ in a tiny dip of the correlation function is thus likely not physical, since the correlation function is mostly flat. (see Fig.~\ref{fig:sum_contributions}).
As a result, %
the total rate contains contributions from the mixed and nuclear-nuclear terms, which leads to a result that is almost but not exactly equivalent to that of GR instanton theory even in cases of small $\Delta$. %
The GR limit would be rigorously recovered by ignoring the tiny dip and taking $x_\alpha\rightarrow\infty$ instead.
\begin{table*}[ht]
\caption{Rate constants $k$ (which have not been normalized by the reactant partition function) for different couplings $\Delta$ for the system of two coupled harmonic oscillators.}
\label{tab:rates}
\centering
\begin{tabular}{l|lllllll}
$\Delta$	& $x_\alpha$ & $k_\text{NA-inst} Z_\mathrm{R}$	& $k_\text{BO-Inst} Z_\mathrm{R}$     & $k_\text{GR-Inst} Z_\mathrm{R}$     & $k_\text{exact} Z_\mathrm{R}$    & $k_\text{Eyring} Z_\mathrm{R}$      & $k_\text{Marcus} Z_\mathrm{R}$ \\
\hline
$10^{-4}$	& 0.83		   & 1.77 $\times 10^{-22}$	& -	                 & 1.86 $\times 10^{-22}$ & 1.95 $\times 10^{-22}$ & - 			    & 1.82 $\times 10^{-30}$ \\ %
$10^{-3}$	& 0.63		   & 1.77 $\times 10^{-20}$     & -		         & 1.86 $\times 10^{-20}$ & 1.95 $\times 10^{-20}$ & - 		    	    & 1.82 $\times 10^{-28}$ \\ %
$10^{-2}$	& 0.37	           & 1.77 $\times 10^{-18}$	& -		         & 1.86 $\times 10^{-18}$ & 1.95 $\times 10^{-18}$ & - 			    & 1.82 $\times 10^{-26}$ \\ %
0.1	        & 0.20		   & 1.79 $\times 10^{-16}$	& -		         & 1.86 $\times 10^{-16}$ & 1.96 $\times 10^{-16}$ & - 			    & 1.82 $\times 10^{-24}$ \\ %
0.25	        & 0.03		   & 1.16 $\times 10^{-15}$     & 1.06 $\times 10^{-13}$ & 1.16 $\times 10^{-15}$ & 1.25 $\times 10^{-15}$ & 1.69 $\times 10^{-22}$ & 1.14 $\times 10^{-23}$ \\ %
0.5	        & 0.0		   & 5.07 $\times 10^{-15}$	& 1.05 $\times 10^{-13}$ & 4.65 $\times 10^{-15}$ & 5.48 $\times 10^{-15}$ & 7.59 $\times 10^{-22}$ & 4.56 $\times 10^{-23}$ \\ %
1.5	        & 0.0		   & 1.16 $\times 10^{-13}$	& 3.99 $\times 10^{-13}$ & 4.19 $\times 10^{-14}$ & 1.27 $\times 10^{-13}$ & 3.06 $\times 10^{-22}$ & 4.10 $\times 10^{-22}$ \\ %
3	        & 0.0		   & 1.01 $\times 10^{-11}$     & 1.72 $\times 10^{-11}$ & 1.67 $\times 10^{-13}$ & 1.12 $\times 10^{-11}$ & 2.48 $\times 10^{-22}$ & 1.64 $\times 10^{-21}$ %
\end{tabular}
\end{table*}
\FloatBarrier

\section{Nonstandard steepest-descent integrals}
\label{app:mathematical_tricks}
In a general treatment of flux correlation functions we want to solve integrals with Dirac $\delta$-functions and Heaviside step functions.
While the former simplifies the integration the latter requires special treatment.
An example of an integral we might want to solve is
\begin{align}
P_\Theta(x_\alpha) &= \Tr\big[\eu{-\tau \Hop/\hbar}\, \Theta(\hat{x} - x_\alpha) \,
\eu{-(\beta\hbar-\tau)\Hop/\hbar} \big] \nonumber \\
&= \Lambda^{-N} \int \mathrm{d}\mathbf{x} \ \eu{-S(\mathbf{x})/\hbar} \, \Theta(x_{N_\tau} - x_\alpha).  
\label{eq:Pgt}
\end{align}
First we locate the instanton as a stationary point of $S$ without considering the step function.
If $\tilde{x}_{N_\tau} \ge x_\alpha$, 
\begin{align}
&P_\Theta(x_\alpha) \sim\Lambda^{-N} \sqrt{\frac{(2\pi\hbar)^{N-1}}{\text{det}_{N-1}\nabla^2 S}} \, \eu{-S(\tilde{\mathbf{x}})/\hbar}
\nonumber \\ &\times
\int \mathrm{d}x_{N_\tau}  \Theta(x_{N_\tau}-x_\alpha) \exp{\left(-\frac{ (x_{N_\tau} -\tilde{x}_{N_\tau})^2}{2\sigma^2}\right)},
\label{eq:app_int_Theta}
\end{align}
where the determinant is taken over all rows and columns except those corresponding to $x_{N_\tau}$
and $\sigma^2 = \hbar \left(\frac{\mathrm{d}^2 S}{\mathrm{d}x_{N_\tau}^2}\big|_{x_{N_\tau}=\tilde{x}_{N_\tau}} \right)^{-1}$.
Here, we chose a simple uniform approximation which is applicable if the Heaviside step function evaluates to one at the stationary point.
We found that this is the case for all semiclassical trajectories that contribute significantly to the rate.
The total derivative of $S$ with respect to $x_{N_\tau}$ is obtained using the finite-difference method.
We %
use the definition of the error function to solve the remaining integral analytically
\begin{align}
&\int_{L_A}^{L_B} \text{d$x$} \ \eu{-\frac{(x-\mu)^2}{2\sigma^2}} \nonumber \\
&= \frac{\sqrt{2\pi\sigma^2}}{2}\left(\text{erf}\left(\frac{L_B-\mu}{\sqrt{2\sigma^2}}\right) - \text{erf}\left(\frac{L_A-\mu}{\sqrt{2\sigma^2}}\right) \right),
\end{align}
where , $L_A = x_\alpha$, $L_B = \infty$ and $\mu = \tilde{x}_{N_\tau}$ in this particular case.

On the other hand, if the optimisation returns a solution for the coordinate $\tilde{x}_{N_\tau}<x_\alpha$, for which the Heaviside step function evaluates to 0, 
the integral will be dominated by its boundary rather than a local maximum.
While we have not found such integrals to contribute to the final rate estimate, they play a key role when the $\alpha$ parameter is far away from its optimal location.
Since their contribution is large in this regime they steer the $\alpha$ optimisation in the right direction. 
In this case, the $x_{N_\tau}$ coordinate is pinned to the most extreme point at which the Heaviside step function still evaluates to 1 (for Equ.~\eqref{eq:app_int_Theta} this would be $\tilde{x}_{N_\tau} = x_\alpha$)
and the rest of the instanton is optimised under this constraint to give $\tilde{\mathbf{x}}'$, where we define $\mathbf{x}' = \{x_0,\dots,x_{{N_\tau}-1}, x_{{N_\tau}+1},\dots, x_{N-1}\}$.
The effective action is expanded around $\tilde{\mathbf{x}}'$
and $\tilde{x}_{N\tau}$.
We define $\tilde{S} = S(\tilde{x}_{N_\tau}, \tilde{\mathbf{x}}')$, 
\begin{equation}
g = \frac{\partial S(x_{N_\tau}, \mathbf{x}')}{\partial x_{N_\tau}}\Big|_{\tilde{x}_{N_\tau}, \tilde{\mathbf{x}}'},
\end{equation}
and
\begin{equation}
\mathbf{H} =
\left.\begin{pmatrix}
    \frac{\partial^2 S(x_{N_\tau}, \mathbf{x}')}{\partial x_{N_\tau}^2} & \frac{\partial^2 S(x_{N_\tau}, \mathbf{x}')}{\partial x_{N_\tau} \partial\mathbf{x}'} \\
    \frac{\partial^2 S(x_{N_\tau}, \mathbf{x}')}{\partial \mathbf{x}' \partial x_{N_\tau}} & \frac{\partial^2 S(x_{N_\tau}, \mathbf{x}')}{\partial {\mathbf{x}'}\partial\mathbf{x}'}
\end{pmatrix}\right|_{\tilde{x}_{N_\tau}, \tilde{\mathbf{x}}'}.
\end{equation}
Noting that the first derivative in $\mathbf{x}'$ is zero, the effective action can be written
\begin{align}
S(x_{N_\tau}, \mathbf{x}') \sim \tilde{S}
+   g    \Delta x_{N_\tau} + \half \begin{pmatrix}
    \Delta x_{N_\tau}\\
    \Delta\mathbf{x}' %
\end{pmatrix}^T  \hspace{-0.2cm} \mathbf{H}
 \begin{pmatrix}
    \Delta x_{N_\tau} \\
    \Delta\mathbf{x}' %
\end{pmatrix}
\end{align}
with $\Delta x_{N_\tau} = x_{N_\tau} - \tilde{x}_{N_\tau}$ and $\Delta \mathbf{x}' = \mathbf{x}' - \tilde{\mathbf{x}}'$.
This can be further simplified using the Schur complement.\cite{Zhang2005}
This decomposition %
allows us to rewrite the integral with the constrained coordinate $x_{N_\tau}$ as
\begin{align}
&P_\Theta(x_\alpha) \sim \int_{x_\alpha}^{\infty} \text{d$x_{N_\tau}$ } \int_{-\infty}^{\infty} \text{d$\mathbf{x}'$ } \eu{- \tilde{S}/\hbar-g \Delta x_{N_\tau}/\hbar} \nonumber \\
&\times\eu{-\frac{1}{2\hbar} \left(\Delta x_{N_\tau} H_{11} \Delta x_{N_\tau} + \Delta \mathbf{x}' \mathbf{H}_{22} \Delta \mathbf{x}' + \Delta x_{N_\tau} \mathbf{H}_{12} \Delta \mathbf{x}' + \Delta \mathbf{x}' \mathbf{H}_{21} \Delta x_{N_\tau} \right)} \nonumber \\
&=
\sqrt\frac{(2\pi\hbar)^{N-1}}{\det_{N-1}\mathbf{H}_{22}} \,
\eu{-\tilde{S}/\hbar}
\nonumber\\ &\quad\times
\int_{x_\alpha}^{\infty} \text{d$x_{N_\tau}$} %
\eu{-\frac{1}{\hbar} g \Delta x_{N_\tau} -\frac{1}{2\hbar} \Delta x_{N_\tau}({H_{11} - \mathbf{H}_{12} \mathbf{H}_{22}^{-1} \mathbf{H}_{21}}) \Delta x_{N_\tau}} .
\label{eq:app_int_Theta3}
\end{align}
The integral over the constrained coordinate $x_{N_\tau}$ can be solved analytically in terms of the error function.
An example of an integral containing two Heaviside step functions is
\begin{align}
&P_{\Theta, \Theta} (x_\alpha,x_\alpha) = \nonumber \\
&\quad\Lambda^{-N} \int \text{d$\mathbf{x}$} \ \eu{-S(\mathbf{x})/\hbar} \, \Theta(x_0 - x_\alpha) \, \Theta(x_{N_\tau} - x_\alpha).
\label{eq:app_int_Theta2}
\end{align}
This integral can be easily solved using an extension of the method described above. %
Since now two explicit integrations are performed, at least one of them has to be computed numerically.
\section{Nonstandard evaluation of the momentum operator}
\label{app:momentum}
In order to compute a flux correlation function that includes at least one nuclear flux operator we need to evaluate the momentum operator at the dividing surface.
In most cases, we can make a simple approximation to obtain the momentum.
For instance, the momentum operator near bead 0 has the effect of multiplication by 
\begin{equation}
p_0 = \frac{\partial S}{\partial x_0}\Big|_{\tilde{x}} %
= m\frac{\tilde{x}_0 - \tilde{x}_1}{\beta_N\hbar},
\end{equation}
which is the standard procedure from conventional BO instanton theory.\cite{InstReview}
\begin{figure}
    \centering
    \includegraphics[width=.99\columnwidth]{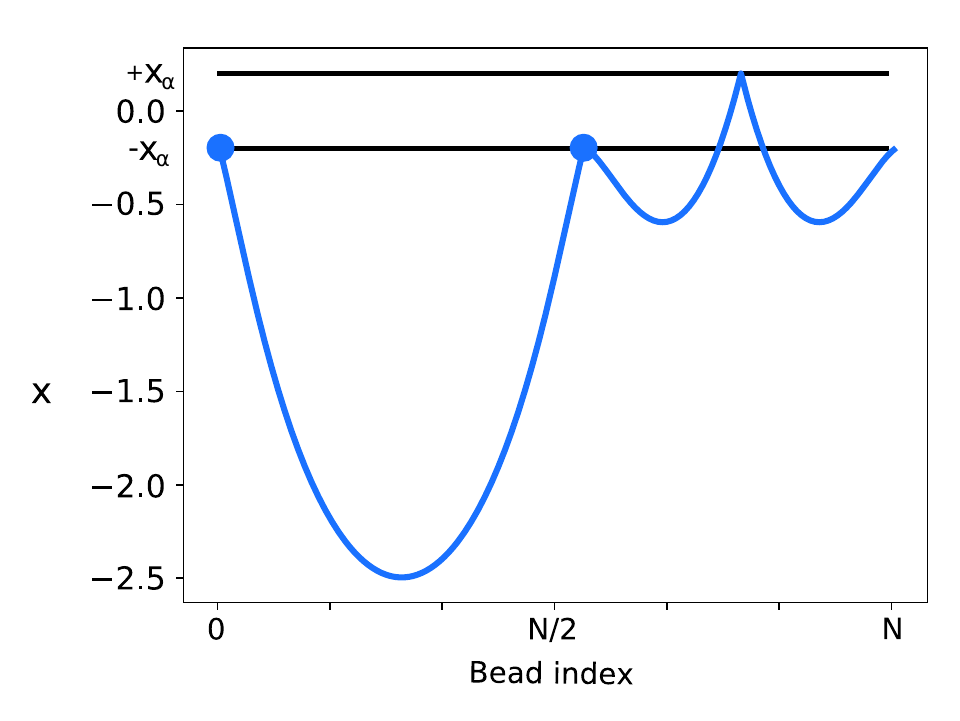}
    \captionsetup{justification=raggedright, singlelinecheck=false}
    \caption{The nuclear coordinates of a 300-bead instanton for the $c^{1, 1, 0, 0}_\mathrm{nn}(0)$ term with $x_\alpha = 0.20$ and a diabatic coupling of $\Delta = 0.25$ is shown in blue.
    The $x_{(N+N_\tau)/2}$ coordinate is pinned to the extremal point $x_\alpha = 0.20$ in order for the corresponding Heaviside step function to evaluate to 1.
    The two dividing surfaces at $\pm x_\alpha$ are drawn in black.
    The location of the two flux operators is highlighted with blue dots.
    }
    \label{fig:mcdonaldsinst}
\end{figure}

A more involved expression for the derivative is required for extreme $x_\alpha$ values, i.e.\ dividing surfaces far from their optimal location. 
In this case, the saddle point falls outside of the Heaviside step function. Therefore, evaluating the flux correlation function with the saddle-point approximation would result in a zero contribution.
Thus, we treat this scenario by Taylor-expanding the action to first order around the boundary of the Heaviside step function.

Let us consider a specific nuclear--nuclear correlation function 
\begin{align}
    c^{1, 1, 0, 0}_\mathrm{nn}(t) = \Tr\big[\Fn{1}\Pe{1} \hat{K}' \Proj^{0}_\mathrm{R} \hat{K}'\Fn{1} \Pe{1} \hat{K}'' \Proj^{0}_\mathrm{P} \hat{K}'' \big].
\end{align}
For the choice of the dividing surface indicated in Fig.~\ref{fig:mcdonaldsinst}, it is necessary to pin one of the beads to the value of $+x_\alpha$ corresponding to the boundary of the Heaviside step function from $\Proj^0_\mathrm{P}$ (see Equ.~\eqref{eq:general_Pp}).      
Plugging in the definition of the flux operators as given in Equ.~\eqref{eq:nucflux}, this can be expanded out to read
\begin{align}
&c^{1, 1, 0, 0}_\mathrm{nn}(t) = 
\frac{1}{4m^2}\Big( \nonumber \\
&\hspace{0.45cm}\Tr\big[\delta(\hat{x} + x_\alpha) \hat{p} \, \Pe{1} \hat{K}' \Proj^{0}_\mathrm{R} \hat{K}' \hat{p} \, \delta(\hat{x} + x_\alpha) \Pe{1} \hat{K}'' \Proj^{0}_\mathrm{P} \hat{K}'' \big] \nonumber \\
&+ \Tr\big[\delta(\hat{x} + x_\alpha) \hat{p} \, \Pe{1} \hat{K}' \Proj^{0}_\mathrm{R} \hat{K}'  \delta(\hat{x} + x_\alpha) \hat{p} \, \Pe{1} \hat{K}'' \Proj^{0}_\mathrm{P} \hat{K}'' \big] \nonumber \\
&+ \Tr\big[\hat{p} \, \delta(\hat{x} + x_\alpha) \Pe{1} \hat{K}' \Proj^{0}_\mathrm{R} \hat{K}'  \hat{p} \, \delta(\hat{x} + x_\alpha) \Pe{1} \hat{K}'' \Proj^{0}_\mathrm{P} \hat{K}'' \big] \nonumber \\
&+ \Tr\big[\hat{p} \, \delta(\hat{x} + x_\alpha) \Pe{1} \hat{K}' \Proj^{0}_\mathrm{R} \hat{K}'  \delta(\hat{x} + x_\alpha) \hat{p} \, \Pe{1} \hat{K}'' \Proj^{0}_\mathrm{P} \hat{K}'' \big]\Big).
\end{align}
As each of the terms in this expansion is evaluated similarly, we will only outline the evaluation of the first term as a guide for the other terms.
This is given by
\begin{align}
&\Tr\big[\delta(\hat{x} + x_\alpha) \hat{p} \, \Pe{1} \hat{K}' \Proj^{0}_\mathrm{R} \hat{K}' \hat{p} \, \delta(\hat{x} + x_\alpha) \Pe{1} \hat{K}'' \Proj^{0}_\mathrm{P} \hat{K}'' \big]
\nonumber \\ 
&\sim \hbar^2 \der{}{x_0} \der{}{x_{N_\tau}} I_1 I_2. \label{eq:nucnuc0001expansion}
\end{align}
In Equ.~\eqref{eq:nucnuc0001expansion}, we introduce two integrals $I_1$ and $I_2$.
In $I_1$ we integrate over the ``special coordinates'', i.e. those to which one applies an electronic and nuclear projection.
The second integral $I_2$ is the steepest-descent integral over the remaining nuclear coordinates $\mathbf{x}' = \{x_1, ...x_{N_\tau/2-1}, x_{N_\tau/2+1}, ... x_{N_\tau-1}, x_{N_\tau+1}, ... x_{(N_\tau+N)/2-1},$ $ x_{(N_\tau+N)/2+1}, ..., x_{N-1}\}$.
\begin{widetext}
The first integral is given by
\begin{align}
I_1 &= \int_{-\infty}^{x_\alpha} \text{d$x_{N_\tau/2}$} \int_{x_\alpha}^{\infty} \text{d$x_{(N+N_\tau)/2}$} %
\,\eu{- g \Delta x_{(N+N_\tau)/2}/\hbar} \, \eu{-\frac{1}{2\hbar}\Delta x_{N_\tau/2}^2 \left(H_{11} - \mathbf{H}_{12} \mathbf{H}_{22}^{-1} \mathbf{H}_{21} \right)},
\end{align}
\end{widetext}
with $g=\frac{\partial S}{\partial x_{(N+N_\tau)/2}}\big|_{\tilde{x}}$, $H_{11} = \frac{\partial^2 S}{\partial x_{N_\tau/2}^2}$, $\mathbf{H}_{12} = \frac{\partial^2 S}{\partial x_{N_\tau/2} \partial \mathbf{x}'}$, and $\mathbf{H}_{21} = \frac{\partial^2 S}{\partial \mathbf{x}' \partial x_{N_\tau/2}}$.
$\mathbf{H}_{22}$ is a reduced Hessian matrix of all but the special coordinates $x_0, x_{N_\tau/2}, x_{N_\tau}$ and $x_{(N_\tau+N)/2}$.
Note that, as already mentioned earlier, the Taylor expansion in the integral over the $x_{(N+N_\tau)/2}$ coordinate is truncated at first order.
This is the bead is pinned to the boundary of the Heaviside step function and is not a saddle point.
The second integral is defined as 
\begin{align}
I_2 = \eu{- S(\tilde{\mathbf{x}})/\hbar}\int_{-\infty}^{\infty} \text{d$\mathbf{x}'$} \eu{-\frac{1}{2\hbar} \Delta\mathbf{x}' \mathbf{H}_{22} \Delta\mathbf{x}'}.
\end{align}
Then, applying the derivatives as given in Equ.~\eqref{eq:nucnuc0001expansion}
\begin{widetext}
\begin{align}
    \hbar^2 \der{}{x_0} \der{}{x_{N_\tau}} I_1 I_2
    \sim \der{S}{x_0} \der{S}{x_{N_\tau}} I_1 I_2
    - \hbar\der{S}{x_0} \der{I_1}{x_{N_\tau}} I_2
    - \hbar\der{S}{x_{N_\tau}} \der{I_1}{x_0} I_2
    - \hbar\ders{S}{x_0}{x_{N_\tau}} I_1 I_2
    + \hbar^2\ders{I_1}{x_0}{x_{N_\tau}} I_2
\end{align}
N.B.\ we have performed the derivative of $I_2$ semiclassically and have treated $\mathbf{H}_{22}$ as slowly varying.
The derivative of $I_1$ is given by
\begin{align}
\hbar \der{I_1}{x_0} \sim -\left(\frac{\mathrm{d} g}{\mathrm{d} x_0} \right) \int_{-\infty}^{x_\alpha} \text{d$x_{N_\tau/2}$} \int_{x_\alpha}^{\infty} \text{d$x_{(N+N_\tau)/2}$} %
\,\Delta_{x_{(N+N_\tau)/2}}\eu{- g \Delta x_{(N+N_\tau)/2}/\hbar-\frac{1}{2\hbar}\Delta x_{N_\tau/2}^2 \left(H_{11} - \mathbf{H}_{12} \mathbf{H}_{22}^{-1} \mathbf{H}_{21} \right)}.
\label{eq:dxI1}
\end{align}
and equivalently for $\rmd I_1/\rmd x_{N_\tau}$.
Similarly,
\begin{align}
\hbar^2\ders{I_1}{x_0} {x_{N_\tau}} \sim\left(\frac{\mathrm{d} g}{\mathrm{d} x_0} \right) \left(\frac{\mathrm{d} g}{\mathrm{d} x_{N_\tau}} \right) \int_{-\infty}^{x_\alpha} \text{d$x_{N_\tau/2}$} \int_{x_\alpha}^{\infty} \text{d$x_{(N+N_\tau)/2}$} %
\,\Delta^2_{x_{(N+N_\tau)/2}}\eu{- g \Delta x_{(N+N_\tau)/2}/\hbar-\frac{1}{2\hbar}\Delta x_{N_\tau/2}^2 \left(H_{11} - \mathbf{H}_{12} \mathbf{H}_{22}^{-1} \mathbf{H}_{21} \right)},
\label{eq:d2xI1}
\end{align}
\end{widetext}
where a second term with $\ders{g}{x_0}{x_{N_\tau}}$ is neglected as it is subdominant. %
Lastly, in Equs.~\eqref{eq:dxI1} and~\eqref{eq:d2xI1} the total derivatives of the gradients of the action are required.
They can be obtained from
\begin{align}
\frac{\mathrm{d} }{\mathrm{d} x_0} \frac{\partial S}{\partial x_{(N+N_{\tau})/2}} = \frac{\partial^2 S}{\partial x_0 \partial x_{(N+N_{\tau})/2}} + \frac{\partial^2 S}{\partial x_0 \partial \mathbf{x}'} \frac{\mathrm{d} \mathbf{x}'}{\mathrm{d} x_0}.
\end{align}
The gradient $\frac{\mathrm{d} \mathbf{x}'}{\mathrm{d} x_0}$ is obtained by solving the linear matrix equation
\begin{align}
\frac{\partial^2 S}{\partial \mathbf{x}'^2}  \frac{\mathrm{d} \mathbf{x}'}{\mathrm{d} x_0} = - \frac{\partial^2 S}{\partial x_0 \partial \mathbf{x}'}.
\end{align}
The derivative with respect to $x_{N_\tau}$ can be obtained equivalently.

\blue{

}

\bibliography{references, references_rhi}
\end{document}